\documentclass[prl,aps,
twocolumn,
tightenlines,
preprintnumbers,
superscriptaddress,
a4paper,nofootinbib]{revtex4-1}
\usepackage{graphicx}
\usepackage{tikz}
\usepackage{epsf}
\usepackage{amsmath,amssymb}             % AMS Math
\usepackage{amsfonts}
\usepackage{slashed}
\usepackage{mathrsfs}
\usepackage{xcolor}
\usepackage{cancel}
\usepackage{nicefrac}
\usepackage{array}

\usepackage{times}
\usepackage{mathptmx}   % Loads the Times-Roman Math Fonts

\newcolumntype{L}[1]{>{\raggedright\let\newline\\\arraybackslash\hspace{0pt}}m{#1}}
\newcolumntype{C}[1]{>{\centering\let\newline\\\arraybackslash\hspace{0pt}}m{#1}}
\newcolumntype{R}[1]{>{\raggedleft\let\newline\\\arraybackslash\hspace{0pt}}m{#1}}

\usepackage[colorlinks,citecolor=blue,linktoc=all,linkcolor=blue,urlcolor=blue]{hyperref} 

\def\beq{\begin{equation}}
\def\eeq{\end{equation}}
\def\bea{\begin{eqnarray}}
\def\eea{\end{eqnarray}}

\def\eqlab#1{\label{eq:#1}}

\def\Eqref#1{Eq.~(\ref{eq:#1})}

\def\Figref#1{Fig.~\ref{fig:#1}}

\def\barr{\left(\begin{array}{c}}
\def\earr{\end{array}\right)}
\def\bmat{\left(\begin{array}{cc}}
\def\emat{\end{array}\right)}
\def\al{\alpha}

\def\si{\sigma}

\def\dd{{\rm d}}

\def\nn{\nonumber}

\begin{document}
%\preprint{MITP/18-xxx}

\title{Comment on ``Measurement of the $Q^2$ Dependence of the Deuteron Spin Structure Function $g_1$ and its Moments at Low $Q^2$ with CLAS
''}

\author{Vadim Lensky}
\affiliation{Institut f\"ur Kernphysik \& Cluster of Excellence PRISMA, Johannes Gutenberg Universit\"at, Mainz D-55099, Germany}
\affiliation{Institute for Theoretical and Experimental Physics, Bol'shaya Cheremushkinskaya 25, 117218 Moscow, Russia}
\affiliation{National Research Nuclear University MEPhI (Moscow Engineering Physics Institute), 115409 Moscow, Russia}

\author{Franziska Hagelstein}
\affiliation{Albert Einstein Center for Fundamental Physics, Institute for Theoretical Physics, University of Bern,  CH-3012 Bern, Switzerland}

\author{Astrid Hiller Blin}
\affiliation{Institut f\"ur Kernphysik \& Cluster of Excellence PRISMA, Johannes Gutenberg Universit\"at, Mainz D-55099, Germany}

\author{Vladimir Pascalutsa}
\affiliation{Institut f\"ur Kernphysik \& Cluster of Excellence PRISMA, Johannes Gutenberg Universit\"at, Mainz D-55099, Germany}

\date{\today}

\begin{abstract}
We argue that the recently published CLAS results on
the deuteron spin polarizability $\gamma_0$ [Adhikari {\it et al.}, 
Phys.\ Rev.\ Lett.\ {\bf 120}, 062501 (2018)], as well as their comparisons with chiral perturbation theory ($\chi$PT),  are misleading. In reality,
the deuteron polarizability is larger by 4 orders of magnitude, as demonstrated here
by a novel calculation in pionless EFT. The CLAS paper, on the other hand, 
presents only a tiny correction to it, based on a partial evaluation
of a sum rule for $\gamma_0$. The sum rule is assumed to have
the same form as for the nucleon; we argue it does not.
Moreover, their ``test of $\chi$PT''  tacitly involves assumptions which, as we demonstrate, may not be valid at the claimed accuracy.  
\end{abstract}

%\pacs{13.60.Fz - Elastic and Compton scattering,
%14.20.Dh - Protons and neutrons,
%25.20.Dc - Photon absorption and scattering,
%11.55.Hx Sum rules}% PACS, the Physics and Astronomy
                             % Classification Scheme.

\maketitle

\begin{figure}[b]
\includegraphics[width=0.85\linewidth]{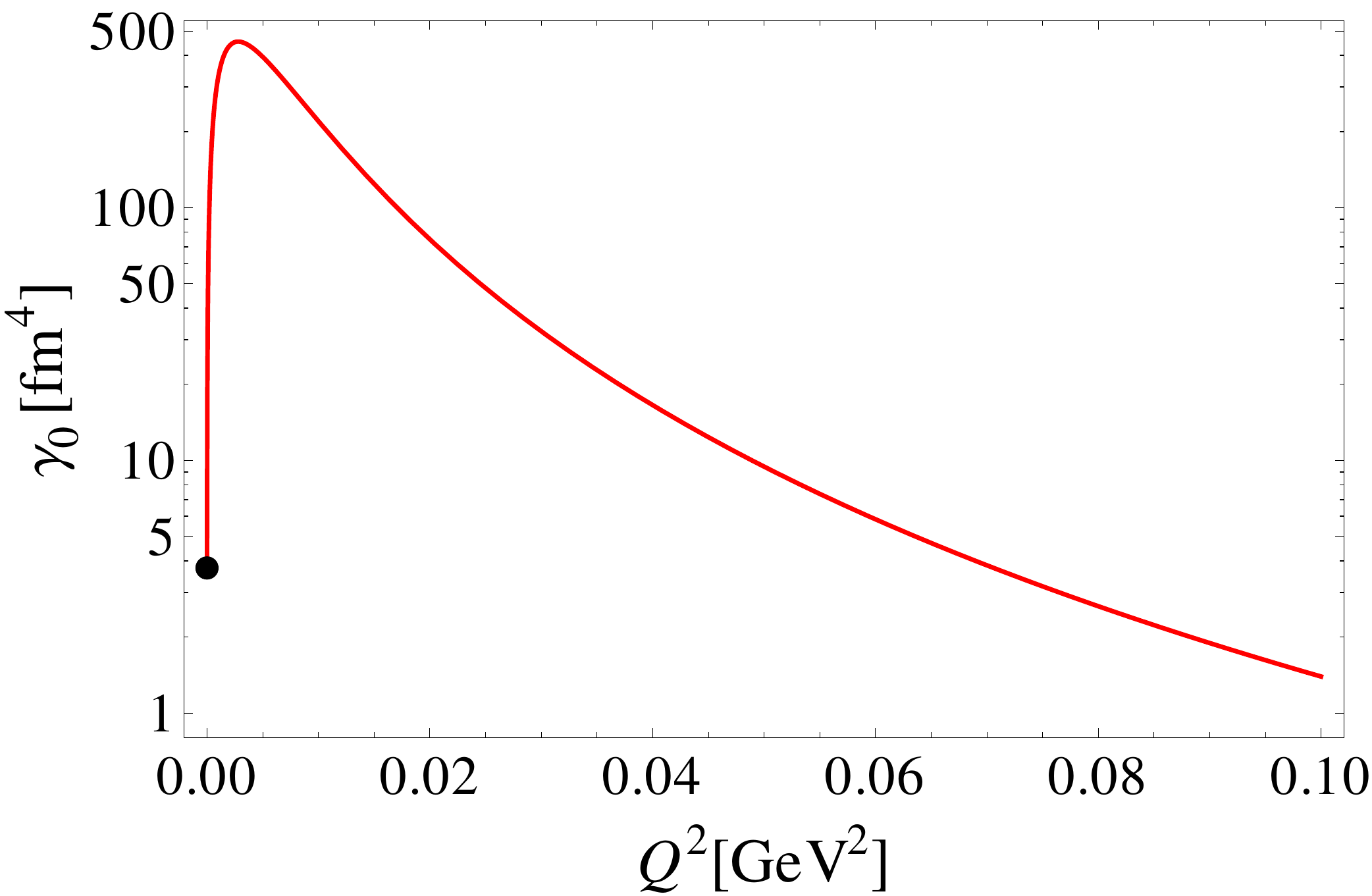}
\caption{Deuteron polarizability $\gamma_0$ calculated in leading-order
$\slashed{\pi}$EFT. The data point at $Q^2=0$ is from \cite{Ahmed:2008zza}.}
\label{fig:LOEFT}
\end{figure}

In a recent Letter \cite{Adhikari:2017wox}, the 
CLAS Collaboration presents   {\it``the first precise measurement of the $Q^2$ evolution of 
$\Gamma_1^d$ and of the spin polarizability $\gamma_0$ on the deuteron''}, shown in their Figs.\ 2 and 3, respectively. 
The results in
Fig.\ 3 are in units of $10^{-4}$ fm$^4$ (cf.\ their arXiv version).
This is 4 orders of magnitude smaller than our estimate,
shown here in \Figref{LOEFT}, based
on extending the pionless-EFT 
calculation~\cite{Ji:2003ia} to finite $Q^2$.

What Fig.\ 3 of \cite{Adhikari:2017wox} really represents is a small contribution to 
the deuteron polarizability, obtained by a partial  evaluation of the 
sum rule in their Eq.~(3).
The upper limit of integration therein is the lowest inelastic threshold
--- in this case, the threshold of the deuteron breakup 
($\gamma^\ast d\to p n$). In the evaluation, however, the
upper limit of integration is set
at a higher energy scale, in the vicinity of the pion-production threshold.
The large low-energy contribution is omitted, without 
stating what the remaining quantity is. 
The same critique concerns the results for $\Gamma_1^d$, albeit the omitted low-energy contribution therein has a much smaller impact because of the different energy weighting.

We stress that \Figref{LOEFT} 
\emph{is} an EFT prediction for the deuteron $\gamma_0$, 
and \emph{not} the ``$\chi$PT results'' presented 
in \cite{Adhikari:2017wox}. The latter are
obtained from the single-nucleon calculations \cite{Bernard:2012hb,Lensky:2014dda}
assuming 
 the deuteron polarizability is given by the isoscalar nucleon polarizability. It is an approximation, which, among
 other effects, neglects the breakup channel. 
 Even though the integral is taken from the pion threshold, where the breakup channel is less important, its effect
 is not necessarily negligible. We have evaluated it at $Q^2=0$, using the
 helicity-difference cross sections of Arenh\"ovel \emph{et al.}~\cite{Arenhovel:2004ha}, with the result of roughly: $-0.7 \times 10^{-4}$
fm$^4$.

Furthermore, Eq.~(3) of \cite{Adhikari:2017wox}
is only correct for a spin-1/2 target, such as the nucleon.
The deuteron has a   
different sum rule \cite{HagelsteinMasters}. 
For $Q^2=0$, it 
reads:
\bea
\eqlab{FSP1}
-\,\frac{ \al }{4M^4}(\varkappa+\varsigma)^2
+2\gamma_0 &=& \frac{1}{4\pi^2}\int _{\nu_0}^\infty \!\dd\nu\,\frac{\si_{0} (\nu) -  \si_{2}(\nu)}{\nu^3 } \\
&=& \lim_{Q^2\to 0} \frac{16\al M^2}{Q^6}\int^{x_0}_0 \dd x\, 
x^2 g_{TT}(x,Q^2), \nn
\eea
with $g_{TT} \equiv g_1 - (2Mx/Q)^2 \, g_2$.  The rhs
(either in terms of  the helicity-difference cross section of
total photoabsorption or, equivalently, in terms of the spin 
structure functions $g_1$ and $g_2$) is the same. The lhs is different.
%\begin{itemize}
%\item[o] 
\emph{The first term}, given by the anomalous magnetic moment
$\varkappa$ and the anomalous quadrupole moment $\varsigma$, 
is absent for the spin-1/2 case. Using
the empirical values for the deuteron:
$ 
\varkappa_{d} \simeq -0.143$, $\varsigma_d \simeq 13.5$, 
$M_d\simeq 1.8756 \; \mbox{GeV}$,
this term amounts to: $- 0.4\times 10^{-4}$
fm$^4$. This would be negligible as a correction to the deuteron polarizability, but not for the small contribution to $\gamma_0$ 
studied in \cite{Adhikari:2017wox}. 
%\item[o] 
\emph{The second term} has a conventional factor of 2, adopted here to compare directly with the literature (e.g., \cite{Ji:2003ia,Ahmed:2008zza}). 
%\end{itemize}

To conclude, 
Ref.\ \cite{Adhikari:2017wox} concerns with only a tiny contribution to 
the deuteron polarizability $\gamma_0$, rather than the polarizability itself. Besides the wrong ``semantics'',
we identify two missing contributions which are potentially 
important: 
1) the
breakup channel, affecting the theory curves, 
in the much discussed comparison  with $\chi$PT; 2) the deuteron 
electromagnetic moments (in general, form factors) in the sum rule of \Eqref{FSP1}.

\vskip2mm

\begin{acknowledgments}
We thank Alexandre Deur for helpful correspondence concerning Ref.~\cite{Adhikari:2017wox}, and 
Hartmuth Arenh\"ovel for kindly providing us with the numerical results of Ref.~\cite{Arenhovel:2004ha}.
This work is supported by the Deutsche Forschungsgemeinschaft (DFG) 
through the Collaborative Research Center [The Low-Energy Frontier of the Standard Model (SFB 1044)], and partly, by the Swiss National Science Foundation. 
\end{acknowledgments}

\end{document}